\documentclass[reprint,showpacs,showkeys,preprintnumbers,amsmath,amssymb,citeautoscript,aip,apl]{revtex4-1}
\usepackage{graphicx}
\usepackage{epstopdf}
\usepackage{color}
\usepackage{float}

\begin{document}

\preprint{Nayak et al.}

\title {Kinetic arrest related to a first-order ferrimagnetic to antiferromagnetic transition in the Heusler compound Mn$_2$PtGa }

\author{Ajaya K. Nayak}
\email{nayak@cpfs.mpg.de}
\affiliation{Max Planck Institute for Chemical Physics of Solids, 01187 Dresden, Germany}
\author{Michael Nicklas}
\affiliation{Max Planck Institute for Chemical Physics of Solids, 01187 Dresden, Germany}

\author{Chandra Shekhar}
\affiliation{Max Planck Institute for Chemical Physics of Solids, 01187 Dresden, Germany}
\author{Claudia Felser}
\affiliation{Max Planck Institute for Chemical Physics of Solids, 01187 Dresden, Germany}
\affiliation{Institut f\"{u}r Anorganische und Analytische Chemie, Johannes Gutenberg
Universit\"{a}t, 55099 Mainz, Germany}%

\date{\today}

\begin{abstract}

 We report a magnetization study of the Heusler compound Mn$_2$PtGa that shows the existence of a magnetic-glass state. Mn$_2$PtGa shows a first-order ferromagnetic (FM)/ferrimagnetic (FI) to antiferromagnetic (AFM) transition in contrast to the martensitic structural transition observed in several Heusler alloys. The kinetic arrest of this first-order FM~(FI) to AFM transition leads to the observed magnetic-glass behavior. We show that the strength of the applied magnetic field, which is the primary parameter to induce the magnetic-glass state, is also responsible for the stability of the supercooled FM (FI) phase in time.
\pacs{}

\end{abstract}

\pacs{}


\maketitle

\section{Introduction}

First-order magnetic to magnetic phase transition possesses a great research interest due to the existence of various anomalous magnetic behaviors. Though there were several previous studies on first-order magnetic phase transition (FOMT), a detailed study was reported in the intermetallic compounds doped CeFe$_2$ \cite{Manekar1,Manekar2} Gd$_5$Ge$_4$ \cite{Levin1}  and in some of the phase separated manganites \cite{Mahendiran1,Hardy1}. Following this, several studies related to FOMT have been reported in other systems such as Nd$_7$Rh$_3$ \cite{Sengupta1} and cobaltites \cite{Sarkar1}. The disordered-broadened first-order transition in most of these compounds arises mainly due to the presence of intrinsic disorder, which serve as nucleation centers below a certain temperature. This first-order transition leads to supercooling, field-induced irreversibility and phase coexistence, that can be eventually termed as a magnetic-glass phase.\cite{Sengupta1,Sarkar1,Roy1} This phase is different from the spin glass one as the glassy state in these systems can be obtained by application of external parameters like magnetic field.

The Ni$_2$Mn$_{1+x}Z_{1-x}$ based Heusler alloys show quite similar types of magnetic properties that of systems showing a magnetic-glass phase. However, these alloys exhibit a structural transition from a high temperature cubic austenite phase to a low temperature tetragonal/orthorhombic martensitic phase with strong magneto-structural coupling.\cite{Krenke1} With help of field cooling it is possible to arrest the high temperature austenite phase below the martensitic transition temperature.\cite{Ito1, Nayak1} It is also observed that one can get large field induced irreversibility in the magnetization loops when measured near to the martensitic transition temperature.\cite{Nayak2,Sharma1} In contrast to the other systems showing magnetic-glass behavior, the field induced structural change in the Heusler alloys plays a major role in inducing the irreversibilities that become stronger on approaching the martensitic transition upon increasing the temperature. However, the other magnetic-glass systems show irreversible behavior at the lowest temperatures that vanishes by increasing the temperature.
In order to explore more functional as well as fundamental properties in the Heusler family, we prepare a new Heusler compound Mn$_2$PtGa \cite{Nayak3}. In this report we show the existence of a magnetic-glass phase in Mn$_2$PtGa that does not show any structural transition. With help of dc-magnetization measurements we show that Mn$_2$PtGa undergoes a first-order ferromagnetic (FM)/ferrimagnetic (FI) to antiferromagnetic (AFM) transition below the magnetic ordering temperature, which is responsible for the observed effect.

\section{Experimental details}

Polycrystalline ingots of Mn$_2$PtGa were prepared by arc melting stoichiometric amounts of the constituent elements in a high purity argon atmosphere. The as-prepared ingots were annealed at 1273 K in an evacuated quartz tube for one week and subsequently quenched in an ice-water mixture. The samples were structurally characterized by x-ray powder diffraction (XRD) using Cu-K$_\alpha$ radiation. Magnetization measurements were carried out on the sample using a superconducting quantum interference device (SQUID) vibrating sample magnetometer (VSM).

\section{EXPERIMENTAL RESULT}

\begin{figure} [b!]
\includegraphics[angle=0,width=7cm,clip]{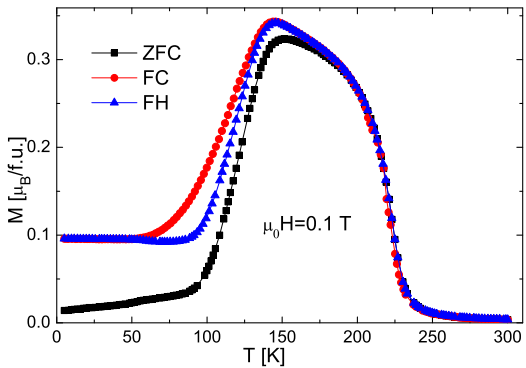}
\caption{\label{FIG1}(Color online)Temperature dependence of magnetization, $M(T)$, measured at 0.1~T.}
\end{figure}


The powder XRD measurement at room temperature reveals that Mn$_2$PtGa crystallizes in a tetragonal structure with space group I-4m2. $M(T)$ measured in zero field cooled (ZFC), field cooled (FC) and field heated (FH) modes is shown in Fig.\,\ref{FIG1}.
In the ZFC mode, the sample was initially cooled to 2\,K and the data were taken upon increasing the
temperature in applied field. In the FC mode, the data were collected while cooling in magnetic field
and subsequently in FH mode the data were collected during heating. All measurements are performed with temperature sweep rate of 3~K/min. It is found that Mn$_2$PtGa undergoes a paramagnetic (PM) to FM~(FI) transition around 230~K followed by a FM~(FI) to AFM transition at 150~K. The small magnetic moment observed in various magnetic measurements suggest that Mn$_2$PtGa orders ferrimagnetically like other Mn$_2$YZ based compounds.\cite{Winterlik11,Klaer11} This is because the Mn atoms occupy two different crystallographic positions with opposite spin alignment. The non-zero nature of the FC/FH magnetization curves below the transition temperature indicates that the low temperature phase is not perfectly AFM in nature. The first-order nature of the low temperature transition is confirmed by the presence of a thermal hysteresis between the FC and FH curves. As the sample shows a tetragonal crystal structure at room temperature, the low temperature transition should not be a structural phase transition, instead a pure magnetic to magnetic one.

\begin{figure} [t!]
\includegraphics[angle=0,width=7cm,clip]{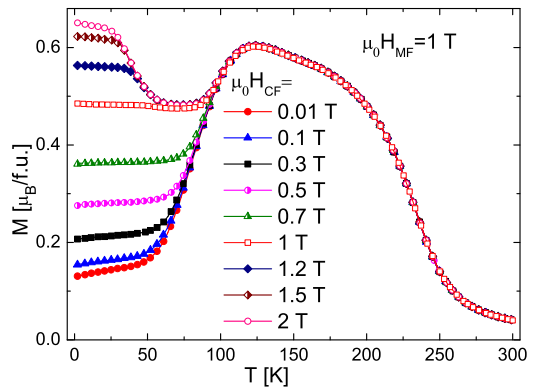}
\caption{\label{FIG2}(Color online) FH $M(T)$ curves measured at 1~T with heating rate of 3~K/min after field cooling the sample in various fields.}
\end{figure}

To probe the existence of a magnetic-glass state in Mn$_2$PtGa, we have performed $M(T)$ measurement in 1~T after cooling the sample in different fields. For this the sample was cooled down to 2~K in presence of a cooling fields $H_{CF}$ from room temperature. At 2~K, the field was changed to the measurement field $H_{MF}$ and the magnetization was measured in heating mode. From Fig.\,\ref{FIG2} two types of magnetic behaviors are observed depending upon the strength of cooling fields. For $H_{CF}\leq$ $H_{MF}$, the $M(T)$ curves show an AFM to FM (FI) transition on increasing temperature. The increase in magnetization with higher cooling fields at low temperatures is due to the increase in the super cooled FM (FI) phase in an AFM background. However, for $H_{CF}>$ $H_{MF}$ the magnetization decreases with increase in temperature followed by an increase around the AFM to FM transition temperature. A higher cooling field will enable a higher amount of supercooled FM (FI) phase. Hence, when the field is reduced to a lower value the system will try to acquire the equilibrium state for that field by a partial conversion of the FM (FI) to AFM phase. This results in an initial decrease in the magnetization at low temperature. It can be mentioned here that irrespective of the strength of cooling field the AFM to FM transition temperature does not change for a fixed measurement field. The above observation confirms that there exists a phase coexistence and a magnetic-glass state in the present sample.

\begin{figure} [b!]
\includegraphics[angle=0,width=7cm,clip]{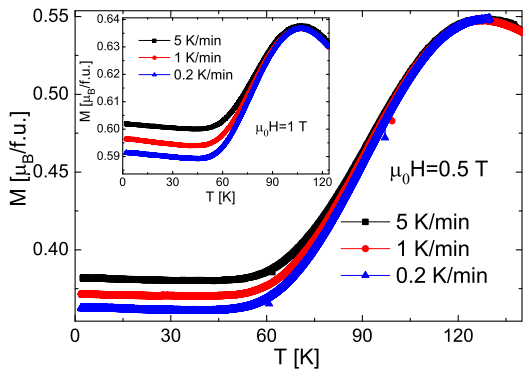}
\caption{\label{FIG3}(Color online) FC $M(T)$ curves measured at 0.5~T with different cooling rates. Inset shows FC $M(T)$ curves measured at 1~T with different cooling rates.}
\end{figure}

For a better knowledge of the stability of the supercooled FM (FI) phase we have studied the temperature sweep-rate dependency of the field cooled magnetization as shown in Fig.\,\ref{FIG3}. For this a field is applied at room temperature and the measurement is performed with a cooling rate of 5 K/min, 1 K/min and 0.2 K/min, respectively. As seen from Fig.\,\ref{FIG3} the magnetization monotonically decreases with decreasing cooling rate. This suggests that with higher cooling rate one can quench the high temperature FM (FI) phase resulting in a higher amount of FM (FI) components in the AFM phase. With a decrease in the cooling rate the supercooled FM (FI) components get a sufficient amount of time to achieve the equilibrium AFM state that results in a lower magnetization value. The process also gives rise to a slight increase in the FM (FI) to AFM transition temperature with lower cooling rate. As both field and cooling rate are responsible for the kinetic arrest of the supercooled FM (FI) phase, it is interesting to see how the magnetization responds to the cooling rate in higher cooling fields. The cooling rate dependence of magnetization measured in a field of 1~T is shown in inset of Fig.\,\ref{FIG3}. It is found that the magnetization changes by 5~\% as the cooling rate changes from 5 K/min to 0.2 K/min in case of a measurement performed in 0.5~T field. Whereas, it is only about 1.7~\% when the measurement is performed in 1~T. As the field stabilizes the FM (FI) phase, with higher cooling field the supercooled FM (FI) phase requires more time to achieve the equilibrium AFM phase. Hence, the effect of the cooling rate becomes less prominent at higher field.

\begin{figure} [tb!]
\includegraphics[angle=0,width=7cm,clip]{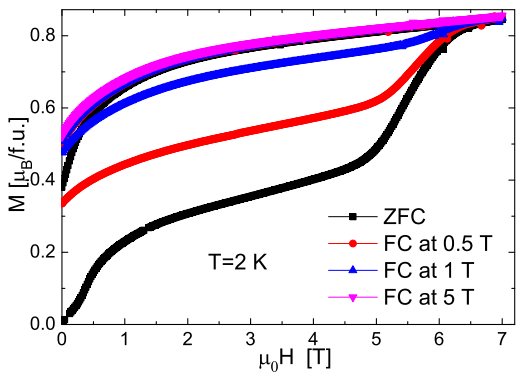}
\caption{\label{FIG4}(Color online) Field dependence of the magnetization, $M(H)$, measured at 2~K with field sweep rate of 0.01 T/sec after field cooling in 0, 0.5~T, 1~T and 5~T.}
\end{figure}

Figure\,\ref{FIG4} shows ZFC and FC $M(H)$ loops measured at 2~K. In case of FC measurements a field is applied at room temperature and subsequently the sample was cooled down to 2~K. It is found that the ZFC $M(H)$ undergoes a field induced metamagnetic transition around 4.8~T. This metamagnetic transition corresponds to the first-order AFM to FM (FI) phase transition. The presence of a metamagnetic transition in the low temperature ZFC $M(H)$ loop confirms the existence of a FM (FI) to AFM transition in Mn$_2$PtGa. To verify the kinetic arrest of the supercooled FM (FI) phase as observed in the $M(T)$ data in Fig.\,\ref{FIG2} and Fig.\,\ref{FIG3}, we have measured the $M(H)$ loops in different field cooled protocols. The $M(H)$ loop measured after FC in 0.5~T shows a higher magnetization value than the ZFC curve. The 1~T and 5~T FC curves follow a similar increasing trend as the 0.5~T FC curve. The critical field for the metamagnetic transition also increases to 5~T and 5.5~T for FC at 0.5~T and 1~T, respectively. No metamagnetic transition can be observed for the $M(H)$ loop measured after field cooling in 5~T. From the above observations it is clear that depending on the strength of the cooling field a certain amount of the high temperature  FM (FI) phase is supercooled at low temperatures that gives rise to larger magnetization value than the ZFC curve. For 5~T field cooling, the sample transfers to the FM (FI) phase in the whole temperature range below the magnetic ordering temperature and hence no metamagnetic transition is observed.

All the above measurements show that the field annealing across the first-order transition forms a new magnetic state at low temperatures. The observation of unequal magnetic behaviors when the sample is cooled with a field smaller/larger than the measuring field clearly indicates the kinetic arrest of the FM (FI) phase. This is further supported by the cooling rate dependence of magnetization measurements, which show that the supercooled FM (FI) phase is highly unstable in time. The stability of the supercooled phase can be increased by higher applied field. The evidence for the FM (FI) phase at the low temperature is also seen from the $M(H)$ loops measured after different field cooling procedure. Furthermore, the present magnetic behaviors are well matched with that observed in various magnetic-glass systems that show a kinetic arrest.\cite{Sarkar1,Roy2} Therefore, it is confirmed that the Mn$_2$PtGa possesses a magnetic-glass phase.

\section{Conclusions}
In conclusion, we have studied the existence of a magnetic-glass phase derived from the first-order FM (FI) to AFM transition in the Heusler compound Mn$_2$PtGa. With help of field annealing across the first-order transition we show that the kinetic arrest of the FM (FI) phase that gives rise to phase coexistence is responsible for the magnetic-glass phase. We also show that the supercooled FM (FI) phase is more stable in time in presence of higher fields.


\begin{acknowledgments}

This work was financially supported by the Deutsche
Forschungsgemeinschaft DFG (Project Nos. TP 1.2-A and
TP 2.3-A of Research Unit FOR 1464 ASPIMATT).
\end{acknowledgments}


\end{document}